\documentclass[useAMS,usenatbib,usegraphicx]{mn2e}
\usepackage{times}
\usepackage{enumerate}

\title[Bayesian photo-z's with empirical training]{Bayesian photometric redshifts with empirical training sets}

\author[C. Wolf et al.]{Christian Wolf \\
 \\
Department of Astrophysics, Denys Wilkinson Building, University of
  Oxford, Keble Road, Oxford, OX1 3RH, UK (email: cwolf@astro.ox.ac.uk).\\
}

\begin{document}
\date{accepted}
\maketitle

\begin{abstract}
We combine in a single framework the two complementary benefits of $\chi^2$-template fits and empirical training sets used e.g. in neural nets: $\chi^2$ is more reliable when its probability density functions (PDFs) are inspected for multiple peaks, while empirical training is more accurate when calibration and priors of query data and training set match. We present a {\it $\chi^2$-empirical} method that derives PDFs from empirical models as a subclass of kernel regression methods, and apply it to the SDSS DR5 sample of $>75,000$ QSOs, which is full of ambiguities. Objects with single-peak PDFs show $<1$\% outliers, rms redshift errors $<0.05$ and vanishing redshift bias. At $z>2.5$, these figures are $2\times$ better. Outliers result purely from the discrete nature and limited size of the model, and rms errors are dominated by the instrinsic variety of object colours. PDFs classed as ambiguous provide accurate probabilities for alternative solutions and thus weights for using both solutions and avoiding needless outliers. E.g., the PDFs predict 78.0\% of the stronger peaks to be correct, which is true for 77.9\% of them. Redshift incompleteness is common in faint spectroscopic surveys and turns into a massive undetectable outlier risk above other performance limitations, but we can quantify residual outlier risks stemming from size and completeness of the model. We propose a {\it matched $\chi^2$-error scale} for noisy data and show that it produces correct error estimates and redshift distributions accurate within Poisson errors. Our method can easily be applied to future large galaxy surveys, which will benefit from the reliability in ambiguity detection and residual risk quantification.
\end{abstract}


\section{Introduction}

Photometric redshifts are an attempt to attribute redshift values to locations in colour space occupied by objects for which we do not have spectroscopic redshifts. Statistically more useful is the aim to attribute expected redshift distributions $n(z)$ to these locations, which are correct in a frequentist interpretation. However, photo-z practitioners are often limited to determine a Bayesian probability distribution $p(z)$, which resembles the state of our knowledge, but differs from the frequentist $n(z)$ by manifestations of ignorance that have to be incorporated to safeguard against known unknowns in the data and in the redshift model.

Photo-z's are obtained using a model expressing the expected colour as a function of redshift and a variation of possible intrinsic colour-affecting parameters. These models come in two distinct flavours with different advantages: (i) template-based models allow the observer to interpret data in empirically unexplored territory by extrapolating the templates in magnitude and redshift space; (ii) empirical models use a subset of the observed objects with independently known redshift, also known as training sets, which are conveniently in the same calibration system as the objects to be estimated. If the training set is truly random, it will also provide the correct priors to the statistical redshift estimation. If it is not, a weights approach as suggested by \cite{Lima08} helps to fix the priors. As a result, smaller or pioneering photo-z surveys have no choice but to use template-based models, while large surveys with small Poisson errors on any of their results wish to control their systematics in the best possible ways and prefer the empirical model, that minimises systematics in the calibration and priors.

After obtaining the data and choosing the model, there remains the choice of estimation code to relate the two. Currently, the two perhaps most advocated categories are $\chi^2$-methods (e.g. \citet{B00}; or \cite{W99} for galaxies and QSOs) and artificial neural nets \citep[ANNs, e.g][]{CL04}. Earlier work has included global or piece-wise polynomial fitting between colours and redshift \citep{Koo85,Con95}. Later the empirical fitting approach has developed into kernel regression methods to optimise local fits \cite[e.g.][]{Wang07,Bo07}; these include nearest-neighbor techniques \citep[e.g.][]{Csa03} and support vector machines \cite[e.g.][]{Wa05}. \citet{Bud09} articulates a unified framework. Here, we note the following general characteristics:

{\it (a) $\chi^2$ model testing} assumes a parametrised model to be free of error by definition and then uses error information on the data to determine probability density functions (PDF) and hence estimates of expectation values and likely errors for the parameters. If the model is correct and error-free, the PDF is expected to be correct, whether the model originates in templates or empirical data.

{\it (b) Kernel regression} uses model realisations with any origin and error properties to estimate a mapping from object features onto parameters and requires smoothing over a local region of the model. When the smoothing (kernel) function is a Gaussian that resembles the data errors, it is identical to $\chi^2$ model testing. 

{\it (c) Conventional ANNs} with a single-number output acting as a parameter estimate deliver unique results. If several parameter values are possible given the same input features, they tend to settle for the most likely one. Errors can be estimated e.g. by resampling the input object as a Gaussian on its error distribution and collecting the outputs into a PDF, which may deliver the possible parameter range of the main solution but might still not help with ambiguities. Probabilistic neural networks (PNNs) which output redshift PDF vectors are currently explored.

The advantage of $\chi^2$ model testing and all PDF-generating techniques is that they evaluate a probability distribution across the range of considered parameters (e.g. redshift) and hence provide a warning signal for ambiguities arising from multiple solutions corresponding to local $\chi^2$ minima. This includes nearest-neighbor techniques that produce PDFs after resampling the input object on its error distribution \citep[as shown for QSOs by][]{Ball08}. In contrast, conventional ANNs deliver the same unique redshift estimate when presented with the same input on different occasions, and thus do not record the relative likelihood of alternative solutions.

Traditionally, practitioners have combined template-based models with $\chi^2$-techniques \citep[starting with][]{Bau62} and empirical training set models with ANNs or kernel regression \citep[e.g.][]{FLS03}. However, the reliability of $\chi^2$-PDFs has been plagued moderately by model deficiencies that could be overcome by using empirical models. This has inspired the following exploration of the $\chi^2$-technique with empirical models, which is an attempt to combine their respective advantages and derive PDFs that are statistically correct and reliable. 

In this paper, we choose to look at photometric redshifts for QSOs in order to confront us with a dataset full of ambiguities (see Sect.~2). We generally use Gaussian kernel functions, and in particular a pure $\chi^2$-empirical approach, described in Sect.~3, on nearly noise-free data. In Sect.~4 we discuss the resulting performance and summarise persistent issues. We look particularly at redshift ambiguities and show how we can use ambiguous objects in further analysis. Sect.~5 aims to give analytic explanations for the origin of redshift error floors, biases and outliers, and supports them with examples from the data. It also provides a framework to evaluate outlier risks in data sets beyond spectroscopic completeness. 

In Sect.~6 we explore the requirements for the $\chi^2$ error scale in the presence of model errors, which allow us to bring the error estimates from the width of the PDF in line with the true redshift errors and to predict how they deviate with different choices of error scale. We note the potentially conflicting interests of optimising a kernel smoothing scale, and propose an approach that combines requirements for smoothing and the statistical error scale in one choice. Finally, using data and a model with different noise levels we demonstrate a reconstruction of a redshift distribution that shows only deviations in line with Poisson uncertainties.

\section{Data}

The purpose of this experiment is to combine the advantages of empirical training samples (calibration and priors implicitly correct) with the advantage of the $\chi^2$-method (ambiguity warning based on a full PDF). We wish to use a data set with plenty of ambiguities to evaluate the benefits of our method.

For most purposes, a large sample of galaxies would be most relevant. However, the only observed galaxy samples large enough for empirical training are provided by SDSS at relatively low redshifts $z\la 0.3$, while strong ambiguities only appear for galaxies at $z>1$. This is why conventional neural networks and nearest-neighbor-techniques have produced extremely robust redshift estimates of the SDSS galaxy sample with precisions of $\sigma_z\approx 0.02$ and virtually no outliers \citep{FLS03,Csa03,Oya08a}. Optical QSO samples are, however, full of redshift ambiguities and thus an ideal testing ground for our purpose.

We opted for the SDSS QSO catalogue by \citet{Sch07}, which is based on SDSS DR5 data but further cleaned and amended. It contains $\sim 77,000$ QSOs ranging in redshift from 0.08 to 5.4, and includes SDSS $ugriz$ photometry as well as 2MASS data for matching objects. A morphology flag marks extended-vs.-unresolved sources ($M=1$ or 0), and where we include it in the $\chi^2$ we assume a fiducial error of $\sigma_M=0.2$ (though this choice makes little difference as $M$ does not carry critical information).

Most objects in the catalogue have vanishing photometric errors on their $ugriz$ measurements, as they are all from a sample which was sufficiently bright for complete spectroscopic follow-up. Exceptions are the bluer bands in $z>2.5$ QSOs, which contain redshifted intergalactic Lyman forest absorption that renders objects fainter and possibly undetected. We found that errors on observed object colours are usually smaller than the intrinsic colour variations exhibited by QSOs at fixed redshift. As a result, our redshift estimation process is limited by the intrinsic properties of the model and not be the data quality. We are thus always in a quality saturation domain, which is appropriate for the study of systematic redshift biases and ambiguities as these are not overshadowed by large statistical errors from low signal-to-noise measurements. Fainter and noisier samples would be additionally affected by wider confidence intervals for the observed photometry and would thus show wider PDFs with larger true and estimated redshift errors.

We clean this catalogue by eliminating objects with missing magnitudes in one or more bands and objects with untypically large photometric errors, eliminating in total 1659 of the 77429 objects ($\sim 2\%$). The remaining sample of 75770 objects is split half and half into a model sample and a data sample, using even-numbered and odd-numbered objects, respectively. The distribution of the two samples is random and statistically similar in terms of magnitude, redshift, and sky position.

\section{Method: empirical $\chi^2$ estimation}

The $\chi^2$-method rests on the following conditions to work properly:

\begin{enumerate}[1.]
\item We compare our data to a model, which needs to represent the possible data appropriately. Using a sample from within the observed data set for the model already ensures that data and model are on a consistent calibration; this is often not the case when external models are used, whether they are template-based or empirical data from an independent project. 

\item The parameter space of the model needs to be broad enough to cover all local minima of the $\chi^2$-distribution, which represent alternative interpretations of the data. Thus, the model sample needs to cover the whole range of parameters expected in the data set; ideally it is a random subset, then the statistical priors will be correct implicitly.

\item In practice the data-model comparison is probed on a discrete grid, which needs to sample the data error distribution properly. Thus, the model needs to be enlarged until its density avoids undersampling; this issue is especially critical for data with ambiguous interpretation: if the technique is expected to be sensitive to a lower-probability secondary solution, this is the one driving the sampling requirements. 

\item We need to know the errors of our data, so that differences between data and model are translated into $\chi^2$ measures and hence probability density functions, while the model is presumed to be error-free.
\end{enumerate}

The empirical $\chi^2$-method is virtually identical to the regularly employed template-based method. The only difference is that we use an empirical set of objects as a discrete model realisation. If we trust that the model is a random subsample of the expected data, then we can use the empirical objects with all the same weight. 

We call $c_{ij}$ the components in the vector of observables for model object $i$. These components could be all the fluxes in different bands, or they could be colour indices, perhaps combined with a single flux value to provide a normalisation. In the case of QSOs, we note that their strong luminosity evolution compensates the dimming with increased distance such that their magnitude distribution hardly depends on redshift below $z\approx 2.5$; at higher redshift, virtually all the redshift constraints are in the colour signature from the Lyman forest. That implies that there is little prior information in their overall brightness, so that colour indices contain basically all the redshift information (and explains why \citet{Ball08} have not found the magnitude priors to be useful).

The probability of a single model object $i$ to give rise to the observed data $c_{{\rm data},j}$ of a given data object is then

\begin{equation}
	p_i \propto \exp^{-1/2 \sum_j [(c_{{\rm model,}{ij}}-c_{{\rm data},j})/\sigma_j]^2}   ~,
\end{equation}

where $\sigma_j$ is essentially a smoothing scale for the weight of the association between a data object and a model object. In a Bayesian framework, we want $\sigma_j$ to be a correct statistical error on $(c_{{\rm model,}{ij}}-c_{{\rm data},j})$ so that $p_i$ expresses the probability of model object $i$ to give rise to the observation of the data object. If the model objects populate the space only sparsely, a smoothing scale is even motivated in a Bayesian framework (see below). 

The expectation value and error estimate for the redshift of the given object follows trivially from the whole model set, after normalising the $p_i$ to $\sum_i p_i=1$:

\begin{eqnarray}
	\langle z_{\rm phot}  \rangle & = & \sum_i p_i \times z_i \\
	\sigma_z			& = & \frac{1}{1+\langle z_{\rm phot}  \rangle} \sqrt{  \frac{1}{n-1} 
					\sum_i p_i (z_i - \langle z_{\rm phot}  \rangle)^2}  ~.
\end{eqnarray}

We use $\sigma_z$ as a redshift quality ranking, and as we show in Sect.~4 objects with low $\sigma_z$ have small true redshift errors as well. 

The full PDF $p_{\rm obj}(z)$ is approximated by the combination of all discrete instances (i.e. objects) in the model. It could be represented e.g. in discrete z-bins after sorting all model objects with weight $p_i$ into the bins for $z_i$. Any shortcomings of this PDF result from the discrete nature and finite size of the model sample. 

Owing to abundant ambiguities in the data, a good fraction of PDFs contain two separate peaks. We
thus test the PDF for bimodality and try to deblend it instead of simply adopting the mean: To this end we split the redshift range into two intervals separated at $\langle z_{\rm phot} \rangle$ and obtain two local solutions $(\langle z_{\rm phot}\rangle,\sigma_z)_{1/2}$. This could already be seen as sufficient if the probability integrals contained in the two different peaks were comparable. But if one peak is a lot less pronounced, the initial mean estimate may lie within the primary mode, so that the $z$-limit between the two intervals splits off and diverts some of the signal from the primary mode into a contamination of the secondary solution.

Hence, we do one more iteration, changing the splitting point to a location in the middle of the two alternative estimates, and redo the estimates again. If the PDF is in fact bimodal, these two solutions represent the two modes just as well as a single mode is represented by the original estimate over the full range. But when a uni-modal PDF is over-deblended with this approach, we find the two resulting estimates to be very close in redshift. We decide in favour of a bimodal PDF  by requiring the redshift difference between the two deblended solutions to be

\begin{equation}
	\frac{1+z_{\rm phot,2}}{1+z_{\rm phot,1}}-1 > 0.4	~.
\end{equation}

This heuristic limit was chosen after visually inspecting a few hundred PDFs and their formal solutions. We note, that this distance requirement corresponds to the width of the redshift interval over which the PDF is integrated for the ODDS parameter in the BPZ code \citep{B00}. When we consider the PDF bimodal, then we flag the object as ambiguous, record the two solutions and determine their relative probability fraction from integrating the PDF over the two ranges. This procedure is sensitive even to ambiguities with a $p$-ratio smaller than $\sim$1-in-20.

The method can be trivially generalised to other object classes, and thus objects can simultaneously be classified on the basis of relative class probabilities and have their class-internal parameters such as redshift estimated within a single framework. This approach has been demonstrated with template-based methods in the COMBO-17 survey \citep{WMR01,W04,W08}. 

If the model is not entirely appropriate for the data, the $p_i$ could be biased away from the correct solution, which may then appear improbable, and thus estimation mistakes might be made confidently. While this is unlikely to happen with empirical models, it can easily result from a choice of inappropriate or incomplete templates or priors, that lead to mismatches between the calibration of data objects and model objects. Even with perfect match, large grid steps can produce discretisation effects when the templates are treated numerically as a discrete grid by the photo-z code.

In a template approach $\sigma_j$ is often enlarged beyond the noise in the data object to include an estimate of essentially unknown but plausible errors in the model using $\sigma_j^2 = \sigma_{{\rm model,}{ij}}^2 + \sigma_{{\rm data},j}^2$. This has been implemented using either constant values providing an error floor on object colours \citep[e.g.][]{WMR01} or template error functions depending on wavelength and template parameters \citep[e.g.][]{EAZY}. As a result, the PDF widens to include the correct solution even when the model is biased, and reduces the rate of catastrophic estimation outliers. However, it should tend to overestimate the statistical redshift errors.

In the following section, we first implement a $\chi^2$-error scale that mimicks a template approach by choosing $\sigma_j =\sigma_{{\rm data},j}$ and pretend the model to be free of errors. Although model biases are not an issue for our empirical model, the presence of model scatter is relevant, but we reserve its rigourous treatment for Sect.~6.

\begin{figure}
\centering
\includegraphics[clip,angle=270,width=\hsize]{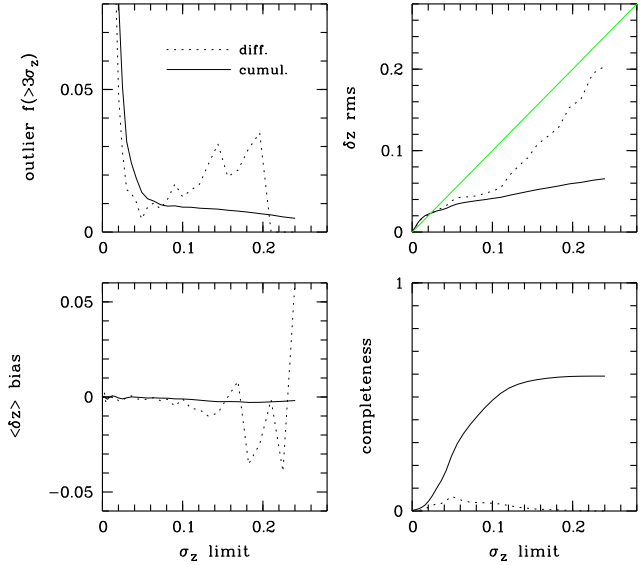}
\caption{Photo-z quality vs. estimated error $\sigma_z$ (non-ambiguous objects only); solid lines show cumulative samples with $\sigma_z<\sigma_{z,\rm limit}$ and dashed lines differential samples at $\sigma_z=\sigma_{z,\rm limit}$.
{\it Top left:} The outlier fraction is generally below 1\%, but it diverges for $\sigma_z \rightarrow 0$ as the tolerance $3\times\sigma_z$ goes to zero as well. 
{\it Top right:} The error estimates $\sigma_z$ are usually larger than the true rms redshift scatter. 
{\it Bottom left:} The bias of the cumulative samples remains within $\pm 0.003$.
{\it Bottom right:} Only 59\% of the objects are classed as unambiguous, and choosing $\sigma_z<0.1$ selects most of them.
\label{stat1}}
\end{figure}

\section{Results using data errors only}

\subsection{Overall performance: RMS, bias and outlier rates}

For general discussion we use the data set with $ugriz$ photometry and the morphology bit, although we briefly comment later on variations that drop the morphology bit or include relatively shallow NIR photometry from 2MASS. We investigate the photo-z quality for a continuous sequence of sub-samples ordered by the expected redshift error $\sigma_z$. Here, we first eliminate all objects flagged as ambiguous ($\sim 41$\%) and discuss them separately in Sect.~4.3 and 5.3. We describe the true photo-z error of each object as

\begin{equation}
	\delta z = \frac{z_{\rm phot}-z_{\rm spec}}{1+z_{\rm spec}}  ~.
\end{equation}

We determine the photo-z quality both for differential samples of objects with expected errors in a small interval around $\sigma_z=\sigma_{z,\rm limit}$ and for cumulative samples of objects with expected errors up to a limit, $\sigma_z<\sigma_{z,\rm limit}$. We characterise the photo-z quality of any sample with the following numbers:

\begin{enumerate}[1.]
\item A fraction of outliers with $|\delta z|>3\times\sigma_{z,\rm limit}$
\item A typical photo-z error, i.e. the rms $\delta z$ of non-outliers
\item A photo-z bias, i.e. the mean $\delta z$ of non-outliers
\item The fraction of the sample with $\sigma_z<\sigma_{z,\rm limit}$ among the full data sample, i.e. the completeness.
\end{enumerate}

The results are presented in Fig.~\ref{stat1}. Outlier rates (top left) are generally below 1\%. The rate goes up as the tolerance goes to zero, just because the true errors remain firmly above zero (at $\sigma_{z,\rm limit}<0.02$ an object with e.g. $|\delta z|=0.06$ is already an outlier). Outliers are more common at $\sigma_z >0.1$ as well, but overall sufficiently rare as to not affect the cumulative samples much.

The true redshift errors (top right) are on average correctly ranked by the estimated errors $\sigma_z$ as shown by the monotony of the dashed line representing the differential sample ordered by $\sigma_z$. Objects with expected $\sigma_z<0.05$ have a true $\delta z$ rms of $<0.05$ as well, but at $\sigma_z>0.05$ errors are overestimated. This is not a desirable statistical property and results from ignoring model errors in the $\chi^2$-empirical approach. We explain this in depth in Sect.~6 and propose an appropriate procedure after discussing the interplay of noise and smoothing scales in producing error estimates.

The photo-z bias (bottom left) is nearly zero for good-quality objects and always $<0.003$ for cumulative samples. Globally, the method is designed to be bias-free, but non-random sub-samples (as the ones plotted here) can always be locally biased. The fraction of objects peaks at $\sigma_z\approx 0.05$, but a cumulative sample must be relaxed to $\sigma_{z,\rm limit}\approx 0.1$ to be $>50$\% complete (bottom right).

In summary, a good-quality subsample selected by $\sigma_z<0.1$ contains half of all objects, shows a bias of $-0.0015$, a $\delta z$ rms of $0.04$ and $0.9$\% outliers. In the following, we investigate the dependence of performance on the desired sample completeness.

\begin{figure*}
\centering
\includegraphics[clip,angle=270,width=0.8\hsize]{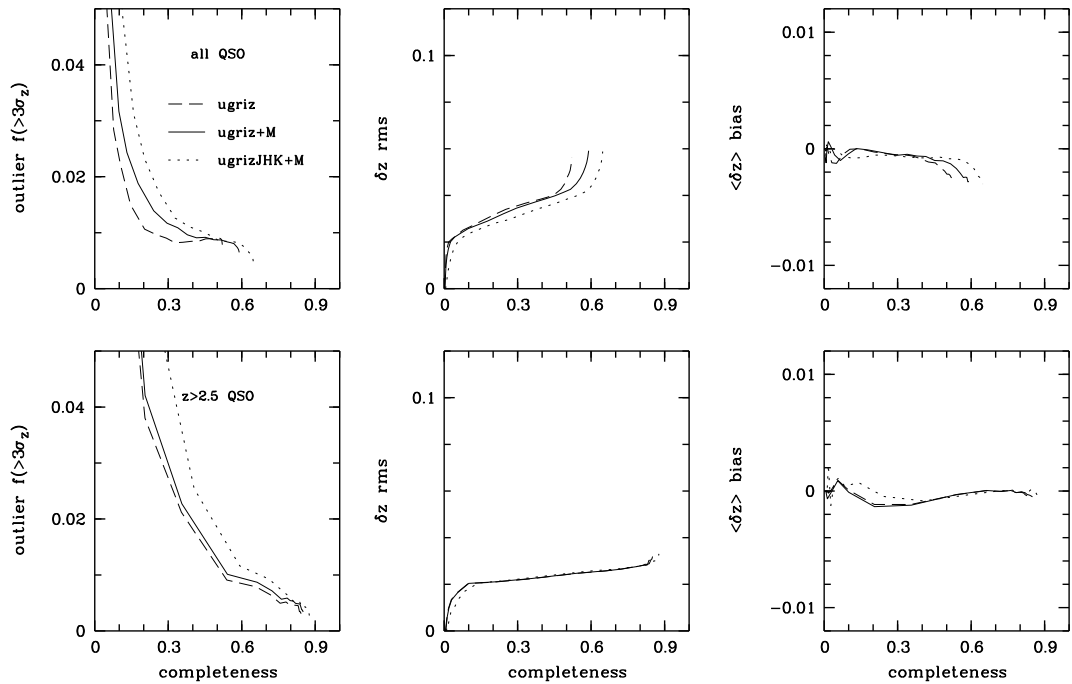}
\caption{Cumulative photo-z quality in dependence of sample completeness: including more objects of worse and worse quality decreases the overall sample quality. {\it Top row:} Adding information on top of the $ugriz$ photometry (2MASS data, morphology; weak constraints only) moves the curves to higher completeness. {\it Bottom row:} QSO photo-z's at $z>2.5$ have low bias and rms values of $\delta z<0.03$. The whole high-redshift sample has only 0.3\% outliers with $\delta z>0.15$. The shallow 2MASS NIR photometry and morphological data do not help in this redshift domain.
\label{qual}}
\end{figure*}

\subsection{Selecting subsamples by estimated photo-z quality}

Many scientific applications are driven by combined requirements for sample size and sample quality. We could thus prefer to choose a quality cutoff for samples from diagnostic diagrams of quality against completeness. Thus, Fig.~\ref{qual} shows the quality numbers over completeness for cumulative samples. We differentiate between overall samples (top row) and high-z only samples (bottom row).

Our default data set is shown by the solid line. The plots confirm that the outlier rate is small for all samples of medium-to-high completeness (top left panel). The rms redshift error remains below 0.04 up to $\sim 50$\% completeness (top centre panel), and only starts shooting up when including the worst 15\% (in terms of $\sigma_z$) of the unambiguous sample. The redshift bias is $<0.003$ for any selection of unambiguous objects (top right panel). All solid lines in the top row end at 59\% completeness as the flagged ambiguous objects are excluded here.

In the high-z sample ($z>2.5$, bottom panels), there are fewer ambiguities expected due to the strength of the Lyman-forest absorption. The completeness reaches 85\% as there are only 15\% detected ambiguities. High-z QSOs can still be confused with low-z objects that have redder SEDs due to substantial host light contribution, but the outlier rate (mostly due to this effect) is less than 1\% at $>55$\% completeness. The cumulative mean redshift error is $<0.03$, and changes little even to the highest completeness. This is simply because at high redshift almost all objects have small $\sigma_z$ and true errors, and completeness increases very quickly with only a little change in $\sigma_{z,{\rm limit}}$. Also, the bias amplitude is $<10^{-3}$ across most of the completeness range.

Finally, two more lines represent alternative data sets: the long-dashed line is obtained when the morphology information is dropped from the $\chi ^2$ analysis, and the dotted line is obtained when shallow JHK photometry from 2MASS is added. The results change little, but generally more information means fewer ambiguities as well as fewer outliers and smaller rms errors. In Fig.~\ref{qual} we would seem to see on the contrary an increasing outlier rate with more information; this is, however, observed at fixed completeness and caused by the outlier tolerance shrinking at fixed completeness alongside smaller error estimates. 

The main effect of added information is to help with breaking ambiguities and thus enlarging the fraction of unambiguous PDFs, i.e. the completeness. Indeed the curves shift to higher completeness with more information, as the fraction of ambiguous objects shrinks. The latter declines from 48\% in the $ugriz$-only data via 41\% when adding the morphology bit to only 35\% after adding shallow $JHK$ data as well. At $z>2.5$ most of the redshift information is contained in the Lyman break, and neither morphology nor shallow NIR data help.

Of course, the fraction of ambiguous objects is expected to collapse dramatically when adding data that really breaks degeneracies such as deep GALEX UV data that helps with $z<2.5$ objects \citep[see e.g.][]{Ball08}, and deeper NIR data that would help with host galaxy light and higher-redshift objects. Adding such data, however, is contrary to the scope of this paper, which is to investigate how our method deals with ambiguities (see next section).

\subsection{Redshift ambiguities}

Redshift ambiguities mean that objects from two or more different redshift regimes appear in the same region of colour space. For a given observed set of colours, two or more redshift solutions are possible, and the most we can know beyond the alternative numbers is their relative probability. The ambiguity can only be broken by adding some discriminating information such as photometry in additional wavebands. Meanwhile, the practical question is how to deal with those ambiguities present. Here, we consider redshift bimodalities as detected by our algorithm, and do not discuss higher-order complications.

If redshift estimates are required for individual objects and statistical redshift distributions for samples are insufficient, there are only two choices: (a) ignore ambiguous objects altogether, or (b) trust the more probable alternative in any bimodal PDF. Our sample contains 15474 ambiguous objects with a mean probability ratio of 78.0\%:22.0\% for the more probable vs. the less probable solution. If these probabilities are statistically meaningful, they ought to represent the success rate of trusting the more probable alternative. In accordance with the estimate, we find for a fact that 77.9\% of ambiguous objects have been attributed to the correct alternative (12051 measured vs. 12077 predicted, well within Poisson noise). We also note, that when the known spectroscopic redshift is used as a prior to choose the right alternative, the error properties are the same as those of unambiguous objects, irrespective of whether we look at the primary or secondary peak.

\begin{figure}
\centering
\includegraphics[clip,angle=270,width=0.75\hsize]{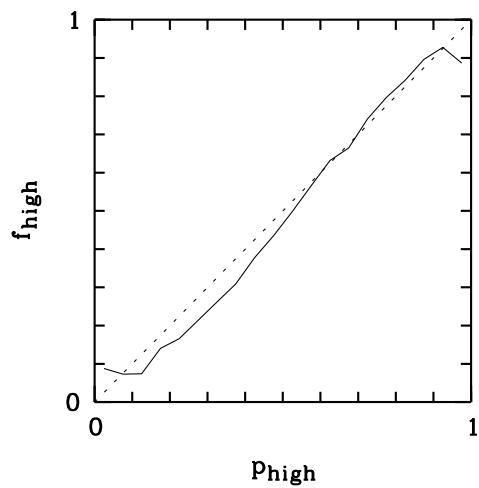}
\caption{Redshift ambiguities are measured roughly statistically correct. The fraction $f_{\rm high}$ of ambiguous objects, where the higher-redshift mode is correct, agrees with the probability fraction $p_{\rm high}$ of this mode in the PDF. 
\label{stat2}}
\end{figure}

Fig.~\ref{stat2} demonstrates that our relative probabilities of ambiguous objects are statistically meaningful even at a subtle level: here, we sort objects into narrow bins of the relative $p$-fraction for the solution at higher redshift, $p_{\rm high}$, which is desired to represent the true fraction of objects $f_{\rm high}$ in this bin, for which the higher-z solution is the correct choice. The figure shows that indeed $f_{\rm high} \approx p_{\rm high}$ everywhere. The relative $p$-fractions thus tell us roughly the risk associated with believing either one of the ambiguous redshift alternatives on an object-by-object basis.

We do not know how sensitive our ambiguity detection really is to $p$-ratios of less than 1-in-20, corresponding to $p_{\rm high}\la 0.05$ or $p_{\rm high}\ga 0.95$. Among our detected ambiguities 8\% have ratios more extreme than 1-in-20, and 1\% even more extreme than 1-in-50. In any case, ambiguities at very low levels exist and will remain mostly undetected, causing catastrophic outliers at the corresponding small rate. If a sample of objects has an ambiguity level of 1-in-50, e.g., they are likely to be classed unambiguous objects at their more probable redshift, and 2\% of them are bound to appear as catastrophic redshift outliers. If e.g. 30\% of an overall sample live in regions of colour space with such a level of ambiguity, this will produce a 0.6\% fraction of unflagged outliers in the overall sample. All existing unflagged outliers in our sample are easily explained by low-level ambiguities remaining undetected.

If trusting risky individual redshifts is unacceptable for the purpose at hand, samples of ambiguous objects can at least be used in a statistical sense, e.g. in the form of redshift distributions $n(z)$ (see following section).

\subsection{Redshift distributions}

Several astrophysical applications do not necessarily require redshifts for individual objects, but can do with redshift distributions for subsamples, e.g. in weak gravitational lensing. Such applications can easily make use of ambiguous objects as well, for which redshift distributions are obtained reliably even though decisions between ambiguous alternatives are risky on an individual basis. The most general solution for any application would be of course to represent any object by its full redshift probability function (PDF) and give up on the concept of a single redshift value. However, we may still continue to use single values occasionally in the interest of data compression.

\begin{figure}
\centering
\includegraphics[clip,angle=270,width=\hsize]{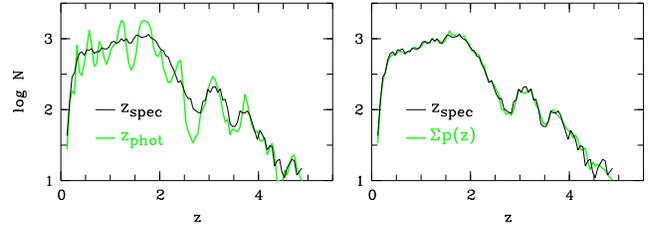}
\caption{Redshift distributions $n(z)$ from spectroscopic and photometric origins are similar. {\it Left:} Objects are represented by $\langle z_{\rm phot} \rangle$; ambiguous objects are counted twice using the two redshifts with their relative probabilities as weights. {\it Right:} Stacking object PDFs matches $n(z)$ much better.
\label{nz}}
\end{figure}

In Fig.~\ref{nz} we look at the summed up distribution of redshifts: comparing spectroscopic redshifts with photo-z's we find good overall agreement. For the photo-z distribution in the ambiguous sample we have counted every object twice, once at each of the alternative redshifts and using the relative probabilities as weights. In the left panel we have represented objects simply by a peak at their estimated redshifts $\langle z_{\rm phot} \rangle$. We expect that true structure on very small scales in redshifts space will be smoothed in photo-z distributions given that photo-z errors mean a lower resolution in observing redshift space. 

However, we find some oscillations in the $n(z_{\rm phot})$, which are not or only weakly present in the $n(z_{\rm spec})$; these have been termed {\it redshift focussing} by some practitioners. One of the reasons for focussing is a local bias arising from $n(z)$ priors inherent in the empirical estimation. Any local maximum in $n(z)$ means that redshifts near the peak are a-priori more probable than in the wings. Objects in the wings then have their overall PDFs biased towards the peak. Peaks are thus overpopulated and troughs in $n(z)$ are depleted.

In the right panel we have used the full $p(z)$ functions of each object and added them up to produce expectation values for the redshift distribution $n(z)$. This is much more similar to the correct redshift distribution and indicates that the PDFs do contain very useful information beyond the redshift expectation value.

\begin{figure*}
\centering
\includegraphics[clip,angle=270,width=0.87\hsize]{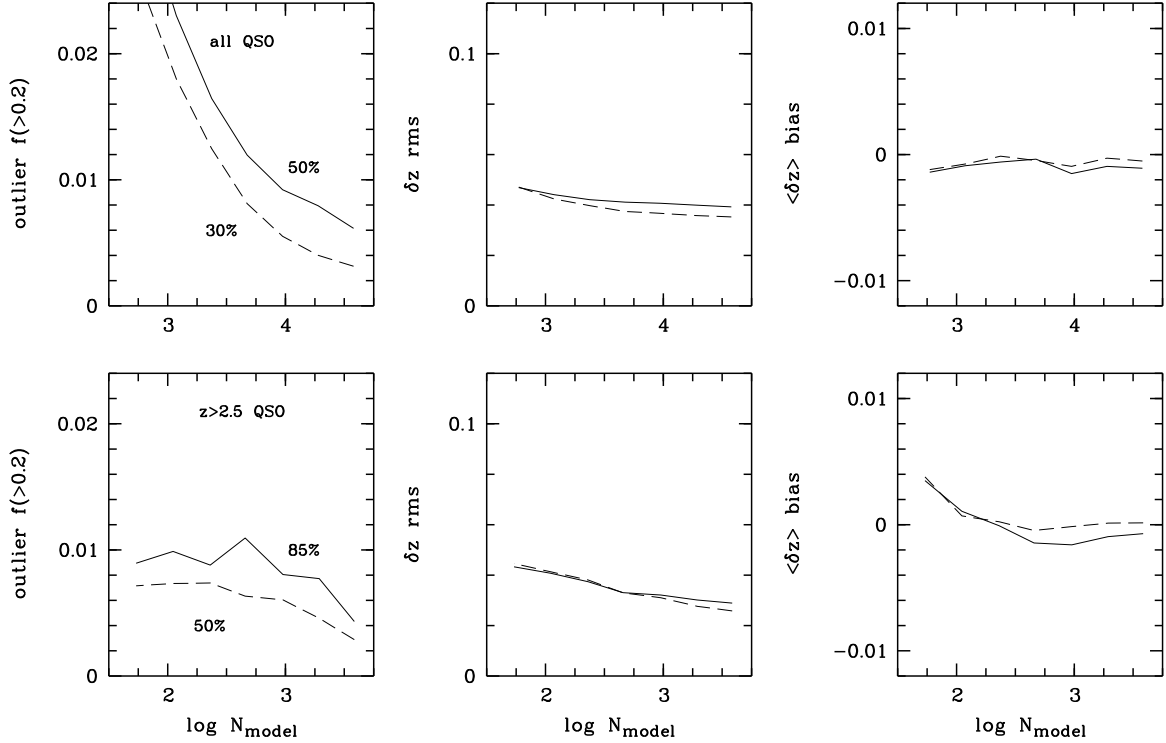}
\caption{Photo-z quality vs. size of the model sample at fixed completeness levels. The quality saturates with ever larger model samples when the photometric error ellipsoids become massively oversampled. The properties of the good-quality bulk of objects (rms redshift errors and bias) do not change much with model size. Outliers (here for fixed tolerance $|\delta z|>0.2$) increase with shrinking model samples whose discrete nature results in loss of sensitivity for redshift ambiguities, except for the high-z sample in which ambiguities are intrinsically rare.
\label{train}}
\end{figure*}

\subsection{Size of model sample}

Generally, the model sample is required to cover the entire range of colours seen in the data sample as otherwise some parts of the latter have no appropriate model to compare with and will be interpreted wrongly. Hence, the volume of colour space covered by the model is fixed, and the size of the model, i.e. the number of its discrete members, then determines its local density in colour space.

We first assume a model distribution without any ambiguities, which is thus not prone to catastrophic outliers. The local density of model objects in colour space now decides how well the error distribution of a data point is sampled. Undersampling can lead to

\begin{itemize}
\item missing the peak of the PDF
\item a local bias in the most likely redshift and thus redshift aliasing
\item larger redshift errors and 
\item an incorrect estimate of redshift confidence intervals
\end{itemize}

Thus, the density required for optimum performance is such that even for well-measured objects with small photometric error ellipsoids, and even in sparsely populated areas of colour or redshift space, the sampling theorem is fulfilled.

We now assume a model with ambiguities and focus on an affected location, where two branches of the model cross in colour space. Now, both branches are required to sample the error distribution of the given data point. Is the density of model points high enough to render even the less populated branch clearly effective, which is now the limiting factor if the ambiguity is to be detected reliably? In this situation, the model density required is a function of the desired sensitivity to ambiguities. Protecting yourself against the rarest possible high-ratio ambiguities requires clearly the most massive model set imaginable.

Fig.~\ref{train} demonstrates the situation quantitatively by reporting the photo-z quality for model samples of different size, where the largest one is the original sample with 37885 objects and the smallest one has only  1/64th of that, i.e. 592 objects. Again, the top row reports results for the entire sample, while the bottom row only reports on high-z objects. The two lines show the photo-z quality at two fixed levels of completeness as a function of $N_{\rm model}$, the number of objects in the model sample.

The most noticable change is a steep increase of outliers when the model shrinks, resulting from ambiguities with extreme $p$-ratios becoming invisible in more sparsely sampled, discrete models when the last object of a more weakly populated branch disappears. In contrast, changes in the bias and rms errors among the non-outlying bulk of data objects are only moderate; these are defined by the intrinsicly densest parts of the model, and are the last statistics to change when the model shrinks to a near-useless size.

\section{Understanding photo-z issues}

We remind the reader that we characterise the performance of photometric redshifts by three parameters at given completeness, all of which we want to optimise, while they keep posing challenges: 

\begin{enumerate}[1.]
\item The rms of the redshift error is supported by the intrinsic scatter of object properties at fixed redshift; enlarging the model sample and getting ever more accurate photometry can not be expected to help in this situation.
\item Sub-samples defined in colour, redshift or quality can have locally biased photo-z's even though the overall sample has vanishing bias by design. 
\item A small fraction of catastrophic outliers can always arise from ambiguities that remain undetected due to large probability ratios between the two alternatives; the only model to guard against all of these is unfortunately just the complete set of spectroscopic redshifts for all objects in the whole survey.
\end{enumerate}

In this section, we try to understand these factors using idealised descriptions of a model and illustrate them with examples from our data set. We hence investigate factors affecting the local properties of photo-z's in small regions of colour space.

\begin{figure}
\centering
\includegraphics[clip,angle=270,width=0.8\hsize]{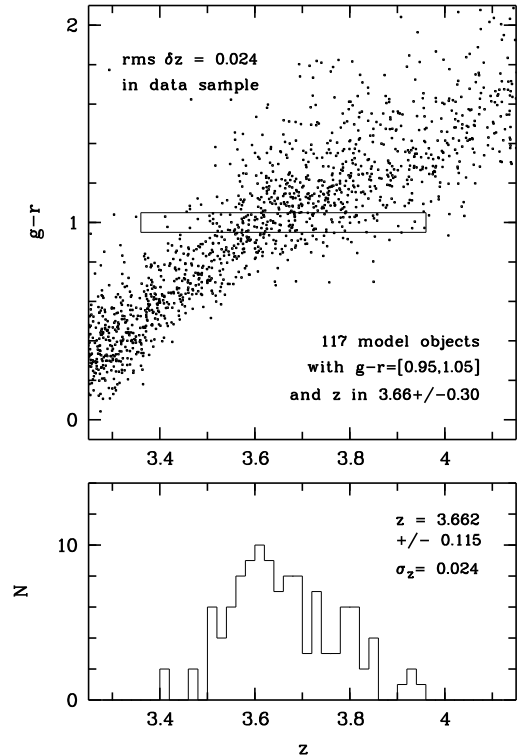}
\caption{Example for the local rms error support from model diversity: At $z\approx 3.7$ the main redshift information is in the $g-r$ colour; at fixed colour, model redshift scatter is the same as the $\delta z$ rms error in data sample. 
\label{scatter}}
\end{figure}

\begin{figure*}
\centering
\includegraphics[clip,angle=270,width=\hsize]{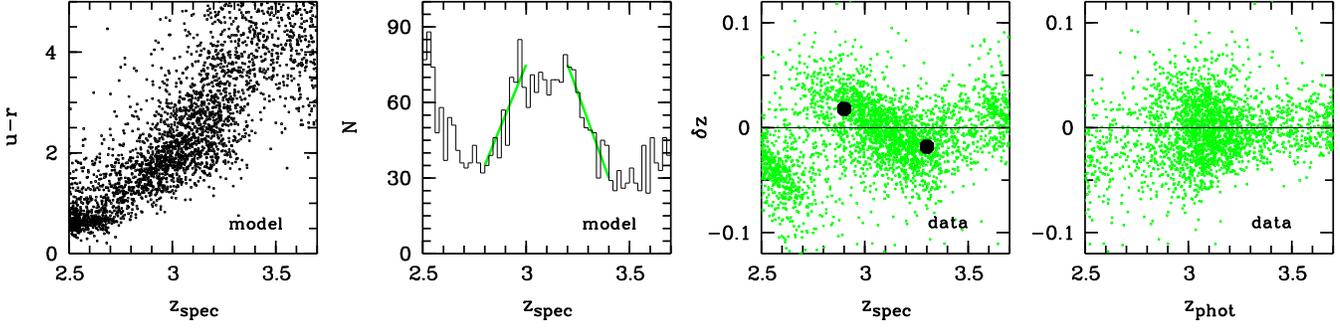}
\caption{Variations in the $N(z)$ of the model (i.e. the priors, centre left panel) cause a local bias of $z_{\rm phot}$ in samples selected by $z_{\rm spec}$ (centre right panel) but no bias of $z_{\rm spec}$ in samples selected by $z_{\rm phot}$ or colour (right panel). See text for more details.
\label{bias}}
\end{figure*}

\subsection{Local RMS error support from intrinsic diversity}

We assume an idealised situation for an analytic approach: an absence of global ambiguities, and a distribution of model and data following two simple rules in a local environment of colour space:

\begin{enumerate}[1.]
\item The mean colour $\langle c(z) \rangle$ of objects at redshift $z$ drifts linearly with $z$ as $\langle c(z) \rangle = c_0 + \beta (z-z_0)$; this applies to both model and data since there is no miscalibration between them.
\item At every $z$ there is a Gaussian distribution in the intrinsic error-free model colours $c(z)$ with an rms $\sigma_{c,\rm model}$ independent of $z$. The errors in the data colours are denoted by $\sigma_{c,\rm data}$.
\end{enumerate}

We can then infer for the model that at fixed colour $c$ there is a Gaussian distribution in $z$ with an rms of $\sigma_{z,\rm model} = \sigma_{c,\rm model}/\beta$, because $dz/dc =1/\beta$. An infinitely accurate colour measurement $c$ in the data will be attributed a PDF with a mean redshift estimate $\langle z(c) \rangle = z_0 + (c-c_0)/\beta$ and an rms of $\sigma_{z,\rm model}$, which is equal to the rms of the true redshift errors found in a data sample.

When a colour error in the data is introduced, the redshift will become less well constrained. Data points of objects at fixed $z$ will then appear with an effective observed colour scatter of $\sigma_{c,\rm eff}^2 = \sigma_{c,\rm data}^2 + \sigma_{c,\rm model}^2$. The mean redshift estimate will be unchanged but the rms of the PDF will widen to 

\begin{equation}
	\sigma_{z,\rm eff}^2 = (\sigma_{c,\rm eff}/\beta)^2 = \frac{\sigma_{c,\rm data}^2}{\beta^2} + \sigma_{z,\rm model}^2	  ~.
\label{sigunbiased}
\end{equation}

Obviously, as long as $\sigma_{c,\rm data} < \sigma_{c,\rm model}$ the intrinsic variety in the model supports a lower bound in the redshift error. The latter will only increase substantially due to noise in the data when it is larger than the model variety.

We now illustrate this point with a concrete example from our data set, and go through the numbers explicitly. We choose objects with a colour of $g-r\approx 1$ that are mostly high-z objects scattered around $z\approx 3.66$. Their main redshift information is in the $g-r$ colour index, which brackets the continuum step over the Lyman-$\alpha$ line due to the intergalactic Lyman-forest absorption, while the other colours provide only weak constraints. We plot the $g-r$ colour of objects near this redshift in Fig.~\ref{scatter} and find a nearly linear colour-redshift relation.

From the data sample, we select objects in a narrow colour interval of $g-r=1.00\pm 0.05$ combined with a redshift interval of $|z-3.66|<0.3$ to exclude rare outlying objects and find 110 unambiguously estimated objects. We find their redshift errors to have a $\delta z$ rms of 0.024 and compare this now to the intrinsic redshift scatter found in the model. From the model, we select 117 objects using the same colour and redshift interval, and expect them to determine the main mode in the PDF of the selected data objects. We find the redshift distribution in the model to have a mean of 3.662 and an rms of 0.115, which translates into $\sigma_z = 0.115/(1+3.662) = 0.024$. The data sample thus shows precisely the errors expected from the redshift scatter inherent to the model in the relevant location of colour space. 

In this example, the photometric noise is too small to contribute to the PDF and error sources. All objects in the data have $g-r$ colour errors below $0\fm04$. We fit a local slope for the redshift-colour relation near $z\approx 3.7$ and find $dz/d(g-r) \approx 0.73$. The propagation of photometric errors alone is thus expected to contribute only a $\delta z$ rms of

\begin{equation}
	\frac{dz}{d(g-r)} \frac{\sigma_{g-r}}{1+z} = 0.73\times 0.04/(1+3.66) \approx 0.006	~. 
\end{equation}

If the data included fainter objects with larger photometric errors, then the colour range of model objects contributing to the PDF would broaden and redshift errors would eventually be driven by the photometric errors: an object with $g-r=1$, e.g., needs $\sigma_{g-r}>0.15$ (i.e. larger than intrinsic scatter) for this to happen.

\begin{figure*}
\centering
\includegraphics[clip,angle=270,width=\hsize]{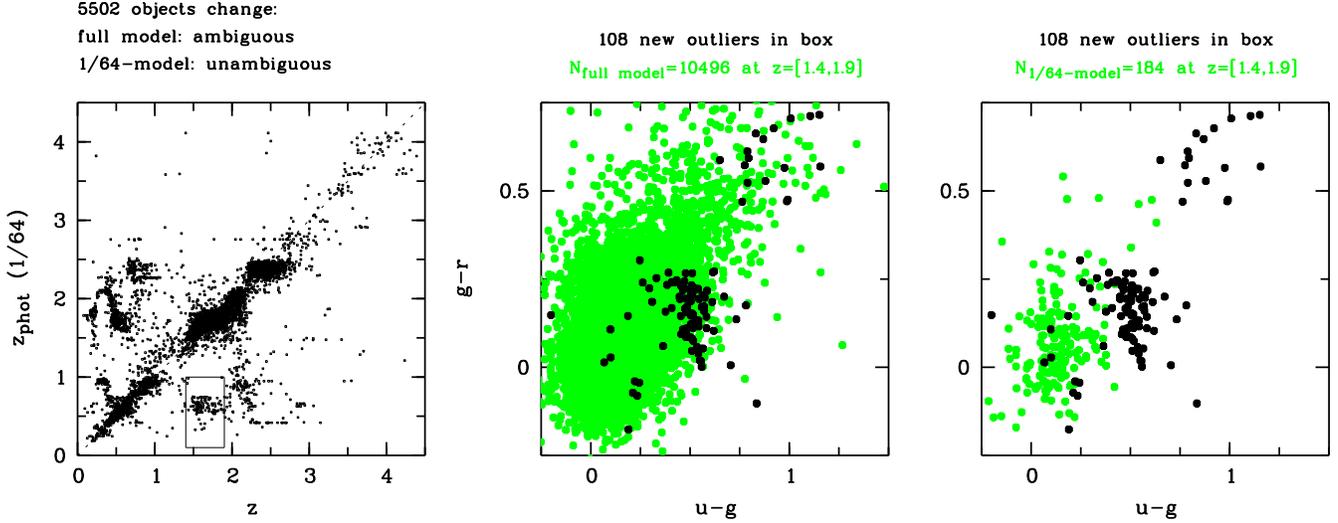}
\caption{Sparse models are less sensitive to ambiguity. {\it Left:} A quarter of all detected ambiguous objects using the full model are classed as unambiguous when the model is stripped to only 1 in every 64 objects, and many of those are then outliers. {\it Centre:} The colours for 108 outliers ($z_{\rm phot}<1$, but true $z=[1.4,1.9]$, see box in left panel) are plotted (black) on top of the plume of colours of the full model at $z=[1.4,1.9]$. {\it Right:} The sparse model does not have much overlap with the 40 objects and suggests a near-zero probability for them to be at $z>1.4$; their ambiguity is not detected any more.
\label{bimodloss}}
\end{figure*}

\subsection{Local redshift biases}

We continue to use the simple approximation from the previous section and introduce a trend of $N(z)$ in the model sample which will act as a prior; we assume a locally constant gradient $dN/dz$ in an environment of $z_0$ with a normalisation of $N(z_0)=1$. A brief calculation including this extension shows that the resulting PDF is skewed by the non-flat prior, the redshift estimate is biased and the estimated error shrinks. The expectation value moves by

\begin{equation}
	\langle z_{\rm eff} \rangle - z_0 = \frac{dN}{dz} \left( \frac{\sigma_{c,\rm data}^2}{\beta^2} + \sigma_{z,\rm model}^2 \right) = \frac{dN}{dz} \sigma_{z,\rm eff}^2 
\label{zbiaseq}
\end{equation}

where $\sigma_{z,\rm eff}$ is given by Eqn.~\ref{sigunbiased}, and the formal rms is now

\begin{eqnarray}
	\langle \sigma_{z,\rm eff, biased}^2 \rangle & = & \sigma_{z,\rm eff}^2 - \left( \frac{dN}{dz} \right)^2 \sigma_{z,\rm eff}^4	~.
\end{eqnarray}

We illustrate this point with an example from a region where $N(z)$ changes strongly with $z$ in the model: Fig.~\ref{bias} shows a region near $z\approx 3$ with a nearly linear colour-redshift relation (left panel), where $N(z)$ changes by a factor of $\sim 2$ in the space of $\Delta z<0.2$, equivalent to $dN/dz\approx \pm 5$ (or $\sim 20$ if expressed on a $z/(1+z)$ scale, see centre left panel). At $z\approx 2.9$ a positive gradient leads to an overestimation of $z_{\rm phot}$ and at $z\approx 3.3$ a negative one leads to an underestimation (see centre right panel). According to Eqn.~\ref{zbiaseq}, we expect average biases in $\delta z$ of $\pm 0.02$ at these two redshift points (given $\sigma_z\approx 0.03$), which is roughly what we find in the data: at $z=2.9\pm0.05$ the bias is $\langle \delta z\rangle =+0.016$ and at $z=3.3\pm0.05$ it is $\langle \delta z\rangle =-0.018$ (the use of a $z/(1+z)$-scale vs. a $z$-scale may be confusing; all quantities need to be on the same scale).

Finally, we want to emphasise that such a bias is indeed not particularly relevant for most photo-z applications as it is not observed, when $\delta z$ is plotted over $z_{\rm phot}$ instead of $z_{\rm spec}$ (see right panel of Fig.~\ref{bias}). By design, the application of priors produces the correct $N(z_{\rm spec})$ for slices in $z_{\rm phot}$- or colour space, although the reverse is not true as exemplified above.

\subsection{Local undetected ambiguities}

The origin of outliers not flagged as ambiguous is a deficiency of the model. We try to give an order-of-magnitude estimate for this residual outlier risks. If we use the largest available model sample, these estimates quantify a maximum risk, which could only be constrained further with the help from a hypothetical larger sample. If we choose a smaller model sample for the photo-z's on purpose, then we can still derive better estimates of the expected outlier rate from the larger one.

For a given object in the data with colour $c$ and error $\sigma_c$, the PDF is mostly defined by the part of the model samples that resides with the central 2$\sigma$-contours of its error ellipsoid. In order to get any moderately reliable PDF, the model sample needs to have at least some objects in this area; a faint ambiguity, i.e. a secondary branch with much lower object density in the same region of colour space, is detected only if there is at least one object seen from that branch. Furthermore, if we observe zero objects of a secondary population, but assume a-priori that it {\it is present} with a flat prior on a density above zero, the expectation value for its density is $\langle N\rangle =1$ objects from Poissonian statistics. Thus, simply assuming a model with $N_{\rm model,local}$ objects in the central part of an error ellipsoid, the residual risk for an undetected faint ambiguity is 

\begin{equation}
  p_{\rm 2nd} = \frac{1}{N_{\rm model,local}}	~.
\label{p2nd}
\end{equation}

This is the maximum oulier risk attached to any single object in this location of the data space. The risks-per-object (as output by the code) add up whenever defining larger data samples from different regions of colour space; they can not be reduced in Poissonian fashion by increasing the size of the data sample, but instead only by increasing the model sample. 

However, the outlier risk can be reduced by stacking the PDFs of several objects across a wider interval of colour space into a summed redshift distribution $n(z)$, because the number of objects in the model contributing to all these PDFs together increases; in this case, $N_{\rm model,local}$ counts all objects in the area of colour space used for stacking. Of course, as we reduce the Poissonian outlier risk and make the PDF more reliable, we also reduce the redshift resolution as we integrate over a wider range of colour space. We are then choosing trade-offs between resolution and reliability.

In Fig.~\ref{bimodloss} we illustrate the effect by comparing photo-z's obtained using a large model set with those obtained using a model set of only 1/64th the size, which has been created from the large one by sparse but random sub-sampling. Looking at the whole data sample, we find that $\sim 1/3$ of the flagged ambiguous objects are not flagged as ambiguous any more when the sparse model is used. The left panel in Fig.~\ref{bimodloss} shows just these 5502 newly unambiguous objects. The majority of them has still good photo-z estimates as indicated by their location near the diagonal, but about 15\% of them have become outliers as a result of overlooking their ambiguity when using the sparse model.

We now focus onto an area containing 108 such {\it 1/64-outliers} in the range of $z=[1.4,1.9]$ and $z_{\rm phot,1/64}<1$ (see box in left panel). The obvious interpretation is that in the colour space occupied by those 108 (z$>$1.4)-objects, the model is dominated by objects of $z<1$, although there is a secondary branch at $z=[1.4,1.9]$ that is visible only in the full model and diluted into discrete absence in the 1/64-model. Supporting this interpretation, we compare the colours of the 1/64-outliers (black) with the plume of model colours across the range of $z=[1.4,1.9]$. In the centre panel, the full model is seen to cover the 1/64-outliers, hence their PDFs contain a mode within $z=[1.4,1.9]$, and ambiguity is detected. But in the right panel, the sparse model is seen to mostly avoid the objects, which eliminates this mode in the PDFs and makes them outliers. Given the small colour errors in the data a colour mismatch of $>0\fm 1$ is already sufficient to suppress the probability of consistency in the $\chi^2$ comparison.

\subsection{Size and incompleteness in model sample; propagation into outliers and redshift bias}

From the arguments in the previous section it is clear that the size and any incompleteness of the model affects the residual outlier risk. We need to distinguish between two kinds of incompleteness, one from variations in targetting objects for spectroscopy, and one from variations in recovering redshifts from target spectra:

\begin{itemize}
\item Incompleteness in targetting means that relatively fewer, but randomly selected objects in a particular part of colour space are included into the model sample. Thus, the priors are not implicitly correct anymore, but this can be compensated by applying explicit weights to the objects.

\item Incompleteness in successfully recovering redshifts from the spectra of target objects is likely to act non-randomly in redshift. It is usually a result of a spectrograph reaching different depth at different redshifts for different galaxy types, due to variations in the strength of spectral features and in their visibility within the instrumental wavelength range. This will omit an important part of the model sample and wipe out its corresponding representation in the PDFs. We point out, that the incompleteness of the PDF is not reduced by simply enlarging the model set, but only by making it more complete in the redshift recovery rate. Also, if the incompleteness affects parts of redshift space strongly by creating redshift deserts, it will translate directly into a similar fraction of undetectable outliers. 
\end{itemize}

The oulier risk translates into a redshift bias risk when averaging the mean redshift of a subsample containing the outliers. At the bright end, undetected outliers are more likely due to sparse sampling of the colour space as a random model sample will contain few bright objects in line with their natural scarcity. At the faint end, they are more likely the result of instrumentally driven selective redshift incompleteness and can reach dramatically high rates \citep[see][and references therein]{New08}. Also, the raw sizes of data sample and model sample translate into simple Poissonian estimates of the expected errors on the mean redshifts.

The Poissonian error on the mean redshift of a subsample localised in colour space is limited by the error on the mean redshift of the stacked PDF, i.e. the number of model objects and the width of their redshift distribution, as well as by the data sample, whose realisation may deviate according to its size:

\begin{equation}
	\sigma_{\langle z\rangle} = \sigma_{z,\rm PDF} \times \sqrt{\frac{1}{N_{\rm model,local}} + \frac{1}{N_{\rm data,local}} }  ~.
\label{Nrms}
\end{equation}

The propagation of the maximum outlier risk into a maximum bias risk also rests on assumptions of their redshift distribution. The mean redshift error of outlying objects $\langle \delta z_{\rm out}\rangle$ in a local region of colour space may be anywhere between $-1+1/(1+z_{\rm max})$ and $+z_{\rm max}$. The maximum bias risk is then:

\begin{equation}
	|\langle \delta z\rangle| = |\langle \delta z_{\rm out}\rangle| \times \left( \eta_{\rm non-recov} + \frac{1}{N_{\rm model,local}} \right)   ~,
\label{Nout}
\end{equation}

where $\eta_{\rm non-recov}$ is the incompleteness of the spectroscopic redshift recovery, and the second factor quantifies the sensitivity limit to ambiguities. If we approximate example numbers supposing $|\langle \delta z_{\rm out}\rangle|=1$, then a 20\% spectroscopic incompleteness implies a maximum redshift bias risk of $|\langle \delta z\rangle|=0.2$, far above any of the $\delta z$ rms values seen in our work. This is a maximum risk, and better constraints require a better model sample.

\emph{
This implies that spectroscopic incompleteness deserves by far the greatest concern in empirical redshift estimation work. This is not much of an issue for any of the bright SDSS samples, but very important for any empirical photo-z work in magnitude regimes where spectrographs do not provide near-100\% completeness.
}

These are rough estimates for overall samples, but again our code produces risk estimates per object, which allow elimination of uncertain objects from samples entering follow-up analyses. Our method is to put all spectroscopic target objects from the model sample without a reliably recovered redshift into a separate model and evaluate the fraction of the total PDF they account for in each individual object; this share of probability is attributed directly to a residual outlier risk for the object in question.

\subsection{Are there ideal sizes for model samples?}

We quantify the ideal size for a complete ($\eta_{\rm non-recov}=0$) model sample, while for incomplete models the conclusion depend on the selections being made for the follow-up analysis. We assume that a data sample will be partitioned into subsamples with stacked PDFs, or each object will be considered on an individual basis ($N_{\rm data, local}=1$); in either case, the critical factor is the size of the model in the local region of the subsample or the individual error ellipsoid, and our rms tolerance on the mean redshift. The following applies if mean redshift of a sample is the critical figure (as e.g. in gravitational lensing) while some applications (such as BAO measurements) may not depend so much on outliers.

The potential rms error is given by Poissonian precision and residual outlier risks, and these change with different powers of the size of the local model sample (Eqn.~\ref{Nrms} and \ref{Nout}). The two error sources have similar impact if

\begin{equation}
	N_{\rm model,local} = 1/\sigma^2_{z,\rm PDF}		~,
\end{equation}

and their combined error estimate is

\begin{equation}
	\sigma_{\langle z\rangle} = \sqrt{2/N_{\rm model,local}}	~.
\end{equation}

If the model has fewer objects than this, the mean redshift of data subsamples is limited by outlier risks that decline $\propto 1/N$, and if it has more it will be limited by redshift scatter that continues to decline $\propto 1/\sqrt{N}$. If the balance is not ideal in a given model sample and enlarging is not an option, there is the alternative of changing redshift resolution and making subsamples cover different ranges of redshift space. As long as the local colour-redshift relation is vaguely linear, this would change model size with the width of the redshift range, $N_{\rm model,local} \propto \sigma_{z,\rm PDF}$; this changes outlier risks in the opposite direction to Poissonian precision and can be used to rebalance the error sources to an optimal mix. 

If we take model incompleteness into consideration, it dominates the error sources as soon as $\eta_{\rm non-recov}>1/N_{\rm model,local}$. Median redshifts, in contrast, are only weakly affected by outliers.

\section{Using model errors in the $\chi^2$ error scale or smoothing function}

In the previous sections we used a model sample that is almost noise-free in conjunction with a $\chi^2$ approach, which is only perfectly reliable when the model is exactly free of noise. We observed that the error estimates showed deviations from the true rms error in Fig.~\ref{stat1} that we still want to explain.
Also, future applications will rely heavily on redshifts of faint objects with noisy photometry. In this section we clarify the the role of noise for the choice of the $\chi^2$-error scale and its consequences for $n(z)$ estimates, and take into account the requirement for smoothing as well.

Assuming that we use a model with full completeness, the data sample and the model sample are drawn from the same parent distribution $\phi(z,c)$ of objects in colour space. Noise, however, smoothes these distributions into a new density function $p(z|c)$, and may differ between the data and model sample. {\it If} both samples are smoothed to the same degree, their $p(z|c)$ are identical. Then the photo-z PDF of any individual data object at location $c_i$ is simply given by $p(z|c_i)$ as determined from the model. Otherwise, we apply an operation to make the smoothing scales consistent. We could also choose to smooth both data and model to a common larger scale. 

If we were prepared to give up the concept of a PDF for an individual object, we could define regions in colour space and attribute the integrated model properties in the region to the corresponding subsample of data objects distributed over that region. In the following, we differentiate two cases, that of a constant target smoothing scale, and that of one which varies across colour space.

\subsection{Spatially homogeneous target smoothing scales}

A spatially homogeneous target smoothing scale is straightforward to deal with, as any object from the data or model sample needs to be smoothed further by an amount that is trivially determined. Every pair of data-model points can be compared separately:

\begin{enumerate}[1.]
\item If the errors of the model object are smaller than the data error, $\sigma_{\rm model}<\sigma_{\rm data}$, we need to smooth the model object further by $\sigma^2_j=\sigma^2_{\rm data}-\sigma^2_{\rm model}$. This is most easily achieved by replacing the model object with a Gaussian of width $\sigma_j$ and evaluating it at the location of the data object $c_i$. We can thus simply use the $\chi^2$-framework described before and use $\sigma_j$ as the error or smoothing scale. Note, that this smoothing scale is obtained by {\it subtracting} the model error from the data error. In contrast, {\it adding} these errors into the smoothing scale smoothes the model too much compared to the data. In the case of an error-free model we recover the usual $\chi^2$ error scale $\sigma_j = \sigma_{\rm data}$.

\item If the errors are nearly identical, then the smoothing scales are already matched. We find the smoothing scale $\sigma_j \rightarrow 0$ and the number of model objects $N\rightarrow 0$ contributing to the solution with diverging Poisson errors. Discretisation effects always call for a sufficient smoothing scale driven by the density of points in the model. We can choose a larger target smoothing scale for both, or we define $p(z)$ only for subsamples distributed over a region in colour space. Thus, holding on to non-zero model smoothing requires to smooth data points as well. This not only wipes out previously present information, but needs to be carried out rigourously (as in 3.) to obtain an unbiased PDF.

\item If the model error is larger than data error, $\sigma_{\rm model}>\sigma_{\rm data}$, we need to smooth the data object, which we implement by resampling the data object as a Gaussian. Model smoothing is still desired for numerical reasons, and a common target scale for data and model needs to be chosen that is larger than either one. Data smoothing is done by resampling as a Gaussian $G(c_i)$ with width $\sigma^2_{\rm resamp} =\sigma^2_{\rm target}-\sigma^2_{\rm data}$ at many $c_{i,j}$, and model smoothing as before by evaluating a model Gaussian at the $c_{i,j}$ as in (1.), whereby $\sigma^2_j=\sigma^2_{\rm target}-\sigma^2_{\rm model}$.
\end{enumerate}

Having $p(z)$ for individual data objects conserves resolution in the data sample that is lost when a common $p(z)$ is attributed to subsamples. {\it But it requires that either data errors are larger than model errors from the start, or noise be introduced into the data after observation, due to a discrete model asking for smoothing! }

\subsection{Spatially varying target smoothing scales}

Here, we discuss only the option of spatially varying smoothing scales that are identical for the data and model sample from the start. Thus, the error scales are already matched, i.e. $\sigma_j=0$ in the $\chi^2$-expression, hence typically no model object would be found at the location $c_i$ of a data object. Again, the only option is to give up on $p(z)$-expressions for individual objects, and to define instead a volume in colour space, over which data objects are combined into a subsample that has the $n(z)$ of the model within the same volume attributed. However, not having $p(z)$ for individual objects means a reduced resolution for the mapping from colour to $n(z)$.

Anyone desiring to use spatially varying scales which differ between data and model sample, will face the problem of finding the target scale for an object in dependence of its original location before the smoothing due to the present errors. This can only be done as a first-order approximation using a representation of the error scales that is already smoothed by the errors itself.

\subsection{Error propagation through the $\chi^2$-expression and the ideal smoothing scale}

We continue to use the idealised example of a locally linear $c(z)$-relation from Sect.~5.1, only that the model is now considered to have a measurement error as well. Thus at fixed redshift, the model has a Gaussian scatter $\sigma^2_{c,{\rm model}} = \sigma^2_{c,{\rm model-int}}+\sigma^2_{c,{\rm model-err}}$ that results from intrinsic scatter convolved with measurement errors. 

A data sample at true redshift $z$ has a colour scatter $\sigma_{c,{\rm data}} = \sigma^2_{c,{\rm model-int}}+\sigma^2_{c,{\rm data-err}}$ that results from the same intrinsic scatter but convolved with the data measurement errors. It translates into a corresponding rms scatter $\sigma_{\langle z \rangle}$ in the redshift expectation values given the local slope of the $c(z)$-relation:

\begin{eqnarray}
	\sigma^2_{\langle z\rangle}	& = & \left( \sigma^2_{c,{\rm model-int}} + \sigma^2_{c,{\rm data-error}} \right) \times \left( \frac{dz}{dc} \right)^2	~.
\label{datarms}
\end{eqnarray}

This result does not depend on the choice of the $\chi^2$ error scale. Only the width of the PDF, and hence the redshift error estimate $\sigma_z$, depends on the $\chi^2$ error scale, as it is a convolution of the model scatter and the model smoothing scale $\sigma_j$:

\begin{eqnarray}
	\sigma^2_z	& = & \left( \sigma^2_{c,{\rm model-int}} + \sigma^2_{c,{\rm model-error}} + \sigma^2_j \right) \times \left( \frac{dz}{dc} \right)^2	~.
\label{esterr}
\end{eqnarray}

Requiring that the PDF and error estimates are representative of the true rms redshift errors, we ask that $\sigma_{\langle z\rangle} = \sigma_z$, which is fulfilled as in Sect.~6.1 when using

\begin{eqnarray}
	\sigma^2_j	& = & \sigma^2_{c,{\rm data-err}} - \sigma^2_{c,{\rm model-error}} 
\label{sigj}
\end{eqnarray}

for a matched error scale. This implies again that the model objects have smaller errors than the data objects, so that the demands of a non-zero smoothing scale to overcome discretisation effects and of the correct error scale can be met simultaneously. The photometry of the query data set only needs to be as good as the lower limit provided by a desired smoothing scale derived from the density of the model points. Otherwise, we have to uphold the desired smoothing scale by introducing additional colour and redshift scatter into the data by resampling the data objects to a larger matched error scale. This means that the photometry of our original query data set was too good to be useful. Conversely, the level of data errors drives the necessary density of model points to suppress outlier risks as discussed in Sect.~5.

We point out the consequence of adding model and data errors in the $\chi^2$ error scale: using $\sigma^2_j = \sigma^2_{c,{\rm data-err}}+\sigma^2_{c,{\rm model-err}}$, we expect to overestimate the true errors by a factor of

\begin{eqnarray}
	\frac{\sigma_z}{\sigma_{\langle z\rangle}} & = & \sqrt{1+\frac{2\sigma^2_{c,{\rm model-error}}}{\sigma^2_{c,{\rm model-int}} + \sigma^2_{c,{\rm data-error}}} }  ~.
\label{sigerr}
\end{eqnarray}

\begin{figure}
\centering
\includegraphics[clip,angle=270,width=\hsize]{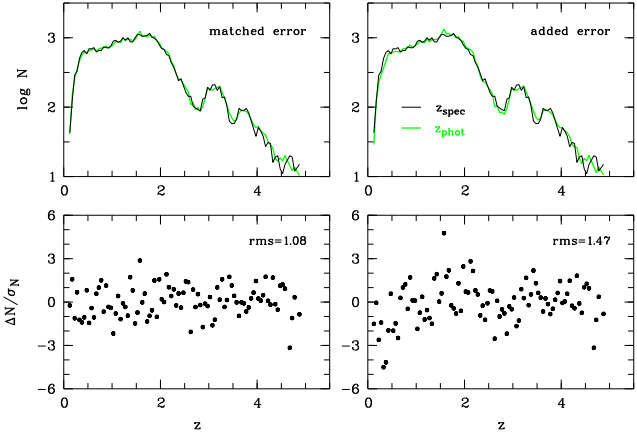}
\caption{Comparison of redshift distributions $n(z_{\rm spec})$ and PDF-stacked $n(z_{\rm phot})$; data and model errors are matched (left) or added (right). The bottom panels show the difference of the $n(z)$ over the expected Poisson error for every redshift bin: Matched errors produce nearly Poissonian $n(z)$ without bias, while added errors produce biases and larger differences.  
\label{nze}}
\end{figure}

A brief numerical experiment is conducted to verify these considerations: we scatter all our data objects to reach a new error of $0\fm1414$ in every colour index, and all our model objects to reach now $0\fm1$. Following Eqn.~\ref{datarms} we expect the matched error scale to produce an rms $\delta z$ error similar to $\langle \sigma_z \rangle$, the mean width of the PDF, while the added error scale should enlarge $\langle \sigma_z \rangle$ by a degree that depends on the intrinsic model scatter and in our case may reach up to a factor of $\sqrt{2}$ where the intrinsic scatter vanishes.

We investigate first the high-z regime, where colour and redshift follow a simple relation. Looking at all objects with $z>3$ but excluding statistical outliers with $|\delta z|>0.1$ or $\sigma_z>0.1$, we measure the mean width of the PDF and the rms $\delta z$ error. In Sect.~5.1 we measured an intrinsic scatter in the model colours of $\sigma_{c,{\rm model-int}}\simeq 0.15$ in this redshift regime, which predicts an increase in $\sigma_z$ by a factor of $\sim 1.2$ using Eqn.~\ref{sigerr}.

The matched error scale produces an rms of $0.0291$ and $\langle \sigma_z \rangle=0.0291$ in perfect agreement. With the added error scale the rms remains almost unchanged at $0.0279$ but the error estimate is increased by a factor of $\sim 1.25$ to $\langle \sigma_z \rangle =0.0366$. We take this as empirical evidence that the added error scale overestimates errors in line with the analytic expectations from the idealised example. 

As a result the stacked PDF from the entire data sample could be wrong near structures in $n(z)$ or edges. In Fig.~\ref{nze} we compare them to the spectroscopic redshift distribution for both error scales. The histogram plots (top row) make it difficult to spot the small differences, but in the bottom row we show the difference between the redshift histograms scaled by the expected Poisson noise in each redshift bin, which is $\sigma^2_N = N^2_{\rm data}+N^2_{\rm model}$ for the difference. The matched error scale shows no apparent bias and an rms scatter of $1.08$ that is very close to Poissonian ($=1$). All bins with 10 or less objects (at the tails of the redshift range) have been eliminated for this plot. In contrast, the added error scale shows drifting biases and an rms scatter roughly enlarged by $\sqrt{2}$.

The conclusion is that the matched error scale approach is an appropriate way to obtain estimates of the redshift distribution which are virtually correct within Poisson noise.

\begin{figure}
\centering
\includegraphics[clip,angle=270,width=\hsize]{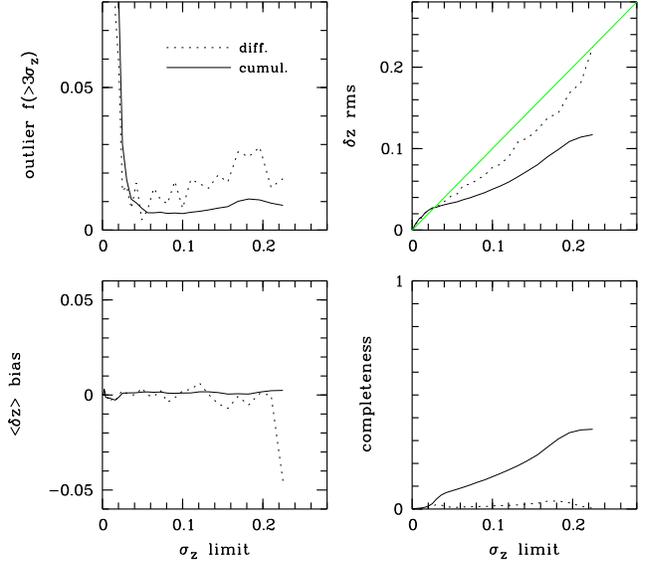}
\caption{Photo-z quality (as in Fig.~1) using noisy data, noisy model and the matched error approach: The error estimates $\sigma_z$ are now comparable to the true rms redshift scatter. Larger noise leads to fewer unambiguous PDFs (completeness). However, bias and outliers remain broadly as expected.
\label{stat1noise}}
\end{figure}

We also repeat in Fig.~\ref{stat1noise} overall photo-z performance figures for the noisy experiment using the matched error scale in the $\chi^2$. We find now that the rms $\delta z$ errors follow roughly the error estimates $\sigma_z$ in the differential sample line, in contrast to the version in Sect.~4 that ignored the model errors. Bias and outlier rates remain as low as before, but the completeness, i.e. the fraction of unambiguous PDFs, has dropped due to the larger errors and smoothing. We revisit the low-noise case in the following section.

\subsection{Revisiting the low-error case}

Armed with the understanding of the impact of smoothing scales onto estimated errors, we reconsider the results of Sect.~4.1, where we decided to use the canonical $\chi^2$-approach while ignoring the model errors. Since the low-noise data were unaltered and randomly split into data and model sample, the two were on a common matched error scale to start with. The application of smoothing to the model only has caused an overestimation of the redshift errors (see Fig.~1), which can now be explained. Given equal errors and our choice of scale $\sigma_{\rm data}=\sigma_{\rm model}=\sigma_j$, we expect to have overestimated errors by up to $\sqrt{2}$; an added error scale could even have led to a factor of up to $\sqrt{3}$, all depending on the relative degree of intrinsic scatter. The alternative of no smoothing suggested by the error scales was of course ruled out by the discretisation effects.

We note, that \citet{Oya08b} estimated errors by smoothing their model with a top-hat kernel function, of course with the motivation to collect enough model objects for a good definition of the PDF. However, they do not find an increase in redshift errors as we do here, and we speculate that this may result from photometric errors being much smaller than the intrinsic colour scatter and the density of their galaxy model sample being extremely high, so that moderate smoothing would introduce only little non-locality and have a very small effect.

\begin{figure}
\centering
\includegraphics[clip,angle=270,width=\hsize]{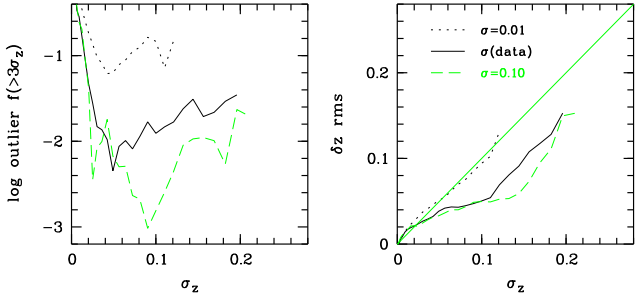}
\caption{Photo-z quality (compare with differential lines in Fig.~1) for low-noise data using different smoothing scales: {\it Right:} The error estimates $\sigma_z$ are nearly identical to the true rms redshift scatter when using near-zero smoothing as suggested by the matched error scale. Larger smoothing overestimates $\sigma_z$. {\it Left:} Smaller smoothing scales lead to larger noise in the PDF and thus higher outlier risks ($\sim10$\% for $\sigma_j=0\fm01$).
\label{stat3sm}}
\end{figure}

Just to prove how the incompatibility of smoothing scales propagates into inappropriate error estimates for the low-noise SDSS QSO data used here, we rerun the experiment from Sect.~4.1 with two fixed scales of $\sigma_j=0\fm01$, close to the desired zero scale and $\sigma_j=0\fm10$, larger than the typical $\sigma_{\rm data}$ levels of $0\fm04$. The results are shown in Fig.~\ref{stat3sm} and compared to the original version using $\sigma_{\rm data}$. It is very clear that the near-zero smoothing produces an excellent correspondence between the $\delta z$ rms and the error estimate $\sigma_z$ (right panel). Larger smoothing scales shift the curve to the right towards progressively overestimated errors.

However, the left panel demonstrates the expected downside of near-zero smoothing, which is a large ($\sim$10\%) fraction of outliers. These appear together with an increased fraction of objects classed to have an unambiguous PDF. A shrinking smoothing function draws its PDF from fewer model objects, overlooks more true ambiguities (see Eqn.~\ref{p2nd}) and produces more residual outliers. In contrast, the large smoothing scale of $0\fm1$ pushed outlier rates below 1\%, even to 0.1\% in parts of the plot.

\subsection{A practical requirement: a constant data error scale}

The results above create a desire for two perhaps conflicting demands: {\it (i)} we want to derive PDFs for individual objects using the matched error scale, which requires a spatially homogeneous target smoothing scale. {\it (ii)} We want to keep smoothing scales on the order of the data errors in order to use the signal contained in the data rather than destroying it with further smoothing. However, if some parts of the data are much noisier than others, they will drive the requirements for the target smoothing scale. Hence, we ideally want to have a constant error across our data sample.

If we are concerned only with bright objects, these may have small errors that may even be dominated by calibration noise and thus be approximately constant on a magnitude scale. On the contrary, when we are concerned with faint objects and diverging magnitude errors, a flux scale is more useful. The errors of faint objects are essentially background noise and constant on a flux scale. Only objects that are brighter than the background have their flux errors growing due to their own Poisson noise or calibration noise. If these are to be treated at the same time as faint objects, a transformed flux scale could be introduced, which maps the mean error as a function of flux onto a constant function. The above procedures could then be exercised using transformed fluxes as object features.

A problem remains even for transformed fluxes, when a large data set includes strong variations in observed depth, as the ideal flux transformation function changes with depth. This challenge is posed by strong variations in interstellar foreground absorption as well. Since object features need to be de-reddened, the intrinsic depth of the data set changes with the absorption level. In these cases, a data set and its model sample may need to be broken down into more homogeneous parts to allow for optimum treatment.

\section{Conclusions}

We have presented a method to obtain Bayesian photometric redshifts using the $\chi^2$-technique with empirical models. This approach is intended to combine in one framework the two complementary benefits of $\chi^2$-template fitting and of empirical training sets as used e.g. by neural networks. The advantage of $\chi^2$-methods is that a probability density function is created, which can be inspected for ambiguities arising from multiple peaks. The advantage of empirical samples is that they can be made to match perfectly the distribution and calibration appropriate for the data sample, as opposed to templates that rely on negotiable assumptions. PDFs generated with imperfect templates can still be unreliable, and where template errors are taken into account in the $\chi^2$, they widen the PDF and increase the error estimates.

Our method produces reliable statistically correct PDFs if a complete empirical model is available. For incomplete models we are able to quantify the mis-estimation risks associated with each individual object. A very simplified description of the comparison could be: Conventional NNs are accurate but unreliable with ambiguities; $\chi^2$-template fitting is less accurate, but guards itself against unreliability with PDFs and template errors; the new $\chi^2$-empirical method is both accurate and reliable.

We used a data set full of ambiguities to demonstrate that the method delivers its promises, i.e. the SDSS DR5 QSO sample with $\sim 75,000$ objects, split half and half into a data and a model sample. Objects with unambiguous PDFs show less than 1\% outliers, typical redshift errors $<0.05$ and vanishing redshift bias. At higher redshift ($z>2.5$) these figures are a factor of $\sim 2$ better. The outliers purely result from the limited size of the model sample, while the rms errors are dominated by the instrinsic variety of QSO colours given the information content in the survey data.

Objects with PDFs classed as ambiguous correctly evaluate the relative probability of the two possible solutions. This provides either accurate weighting factors when using both interpretations for an object in a later analysis, or an accurate outlier risk when using only the more probable solution. In the latter case, our method predicted that in 78.0\% of ambiguous objects the more probable peak in the PDF would be the correct one, which was then found to be true for 77.9\% of them, different by less than Poisson noise.

The method had been inspired by the template-based photo-z code employed in the CADIS and COMBO-17 surveys \citep{W99,WMR01}, except that it replaces template realisations on a grid with empirical model objects. It is thus also capable of classifying objects with Bayesian probabilities into stars, galaxies, QSOs and into various subclasses. 

For noisy data we propose a {\it matched error} approach, which is designed to compare data and model at common resolution in colour space. This translates into a $\chi^2$-error scale given by $\sigma^2_{\rm data}-\sigma^2_{\rm model}$, and we show that this method provides accurate error estimates. In contrast, adding data and model errors in the $\chi^2$-expression broadens the probability distribution and thus overestimates the rms redshift errors. Finally, we show that the matched error scale in the $\chi^2$-empirical method reconstructs the redshift distribution of a noisy data sample practically within Poissonian $n(z)$-errors, if a complete albeit noisy empirical model is available.

The method is most easily implemented when object features can be transformed onto a scale where data errors are constant and model errors are smaller than data errors. Then the model smoothing is provided by matching the error scales in the $\chi^2$ expression and no data resampling is required. In this case, the procedure is computationally very fast; e.g. the QSO sample in this work was processed in 20 minutes on a year 2004 PowerPC Mac laptop.

Empirical models are more representative of the data, and thus the derived PDFs are substantially more accurate than PDFs derived from template fitting, allowing to trust redshifts, ambiguities and outlier risk evaluations, which is critical for understanding systematics in large photo-z data sets. However, their limitations arise principally from the size and completeness of the model sample. Redshift-selective incompleteness as it often appears at the faint end of spectroscopic surveys translates into a massive undetectable outlier risk that can far exceed any of the other performance limitations. While such incompleteness is the main challenge for any empirical method, we provide a framework to evaluate catastrophic risks for individual objects as to allow for their separate handling.

An important application of future photo-z work is in massive cosmological surveys for galaxy photo-z's, which will indeed require superb control of systematics such as redshift biases and outliers. The results presented here for QSOs are not applicable to galaxies in a quantitative sense, but our use of QSOs was motivated by the rich ambiguities present, which for galaxies are only expected in future large samples. When they become available, they will benefit just as well from our method that derives robust Bayesian photo-z's from empirical samples and evaluates residual risks for outlier rates and photo-z biases.

\section*{acknowledgements}
CW was supported by an STFC Advanced Fellowship and is very grateful for  suggestions from an anonymous referee that led to further work and an improved paper.

Funding for the creation and distribution of the SDSS Archive has been provided by the Alfred P. Sloan Foundation, the Participating Institutions, the National Aeronautics and Space Administration, the National Science Foundation, the US Department of Energy, the Japanese Monbukagakusho, and the Max Planck Society. The SDSS Web site is http://www.sdss.org. The SDSS is managed by the Astrophysical Research Consortium for the Participating Institutions. The Participating Institutions are the University of Chicago, Fermilab, the Institute for Advanced Study, the Japan Participation Group, The Johns Hopkins University, the Korean Scientist Group, Los Alamos National Laboratory, the Max Planck Institute for Astronomy, the Max Planck Institute for Astrophysics, New Mexico State University, the University of Pittsburgh, the University of Portsmouth, Princeton University, the United States Naval Observatory, and the University of Washington.


\begin{thebibliography}{}
\bibitem[\protect\citeauthoryear{Ball et al.}{2008}]{Ball08}
  Ball, N. M.,, Brunner, R. J., Myers, A. D., Strand, N. E., Alberts, S. L. \& Tcheng, D.,
  2008, ApJ, 683, 12
\bibitem[\protect\citeauthoryear{Baum}{1962}]{Bau62}
  Baum, W. A., 1962, in {\it Problems of Extragalactic Research},
  ed. McVittie, G. C., Macmillan Press, IAU Symposium 15, 390 
\bibitem[\protect\citeauthoryear{Benitez}{2000}]{B00}
  Benitez, N., 2000, ApJ, 536, 571
\bibitem[\protect\citeauthoryear{Boris et al.}{2007}]{Bo07}
  Boris, N. V., Sodre, L., Jr., Cypriano, E. S., Santos, W. A., Mendes de Oliveira, C. \& West, M.,
  2007, ApJ, 666, 747
\bibitem[\protect\citeauthoryear{Brammer et al.}{2008}]{EAZY}
  Brammer, G. B., van Dokkum, P G., Coppi, P., 2008, ApJ, 686, 1503
\bibitem[\protect\citeauthoryear{Budavari}{2009}]{Bud09}
  Budavari, T., 2009, ApJ, 695, 747
\bibitem[\protect\citeauthoryear{Collister \& Lahav}{2004}]{CL04}
  Collister, A. A. \& Lahav, O., 2004, PASP, 116, 345
\bibitem[\protect\citeauthoryear{Connolly et al.}{1995}]{Con95}
  Connolly, A. J., Csabai, I., Szalay, A. S., Koo, D. C., Kron, R. G. \& Munn, J. A., 1995, AJ, 110, 2655
\bibitem[\protect\citeauthoryear{Csabai et al.}{2003}]{Csa03}
  Csabai, I., Budavari, T., Connolly, A. J., Szalay, A. S., et al., 2003, ApJ, 125, 580
\bibitem[\protect\citeauthoryear{Firth et al.}{2003}]{FLS03}
  Firth, A. E., Lahav, O. \& Somerville, R. S., 2003, MNRAS, 339, 1195
\bibitem[\protect\citeauthoryear{Koo}{1985}]{Koo85}
  Koo, D. C., 1985, AJ, 90, 418 
\bibitem[\protect\citeauthoryear{Lima et al.}{2008}]{Lima08}
  Lima, M., Cunha, C. E., Oyaizu, H., Frieman, J., Lin, H. \& Sheldon, E., 2008, MNRAS, 390, 118
\bibitem[\protect\citeauthoryear{Newman}{2008}]{New08}
  Newman, J. A., 2008, ApJ, 684, 88
\bibitem[\protect\citeauthoryear{Oyaizu et al.}{2008a}]{Oya08a}
  Oyaizu, H., Lima, M., Cunha, C. E., Lin, H., Frieman, J. \& Sheldon, E. S., 2008a, ApJ, 674, 768
\bibitem[\protect\citeauthoryear{Oyaizu et al.}{2008b}]{Oya08b}
  Oyaizu, H., Lima, M., Cunha, C. E., Lin, H. \& Frieman, J., 2008b, ApJ, 689, 709
\bibitem[\protect\citeauthoryear{Schneider et al.}{2007}]{Sch07}
  Schneider, D. P., Hall, P. B., Richards, G. T., Strauss, M. A., Vanden Berk, D. E., et al., 2007,
  ApJ, 134,102
\bibitem[\protect\citeauthoryear{Wadadekar}{2005}]{Wa05}
  Wadadekar, Y., 2005, PASP, 117, 79
\bibitem[\protect\citeauthoryear{Wang et al.}{2007}]{Wang07}
  Wang, D., Zhang, Y. X., Liu, C. \& Zhao, Y. H., 2007, MNRAS, 382, 1601
\bibitem[\protect\citeauthoryear{Wolf et al.}{1999}]{W99}
  Wolf, C., Meisenheimer, K., R\"oser, H.-J., et al., 1999, A\&A, 343, 399  
\bibitem[\protect\citeauthoryear{Wolf et al.}{2001}]{WMR01}
  Wolf, C., Meisenheimer, K., R\"oser, H.-J., 2001, A\&A, 365, 660  
\bibitem[\protect\citeauthoryear{Wolf et al.}{2004}]{W04}
  Wolf, C., Meisenheimer, K., Kleinheinrich, M. et al., 2004, A\&A, 421, 913
\bibitem[\protect\citeauthoryear{Wolf et al.}{2005}]{WGM05}
  Wolf, C., Gray, M., \& Meisenheimer, 2005, A\&A, 443, 435
\bibitem[\protect\citeauthoryear{Wolf et al.}{2008}]{W08}
  Wolf, C., Hildebrandt, H., Taylor, E. N. \& Meisenheimer, K., 2008, A\&A, 492, 933
\end{thebibliography}
\end{document}